\documentclass[preprint,showpacs,preprintnumbers,amsmath,amssymb,aip]{revtex4-1}

\usepackage{graphicx}
\usepackage{dcolumn}
\usepackage{bm}
\usepackage{psfrag}

\begin{document}

\preprint{APS/123-QED}

\title{Bounds on poloidal kinetic energy in plane layer convection}

\author{A. Tilgner}

\affiliation{Institute of Geophysics, University of G\"ottingen,
Friedrich-Hund-Platz 1, 37077 G\"ottingen, Germany }

\date{\today}

\begin{abstract}
A numerical method is presented which conveniently computes upper bounds on heat
transport and poloidal energy in plane layer convection for infinite and finite
Prandtl numbers. The bounds obtained for the heat transport coincide with
earlier results. These bounds imply upper bounds for the poloidal energy which
follow directly from the definitions of dissipation and energy. The same
constraints used for computing upper bounds on the heat transport lead to
improved bounds for the poloidal energy.
\end{abstract}

\maketitle

\section{Introduction}

Transport in turbulent flows, for instance heat flow across a turbulent
convecting layer, cannot be computed analytically but has to be computed
numerically or measured experimentally. However, rigorous analytical
calculations can provide us with upper bounds on the transport. The oldest work
finding upper bounds on turbulent dissipation seems to go back to Hopf
\cite{Hopf41}. Bounds for turbulent heat transport (which can also be expressed
in terms of dissipation) appeared later in different forms. A series of papers
including refs. \onlinecite{Malkus56,Howard63,Busse69} leads to one method for obtaining
upper bounds, called MHB method for short in this paper, and Constantin and Doering
presented another method closely related to Hopf's original work 
\cite{Doerin92,Doerin96} denoted as
CDH method in the sequel. These two methods yield identical bounds if the best
bounds obtainable within each method are compared. More
recently, Seis \cite{Seis15} came up with another approach which
obtains bounds from a very compact derivation.

The quality of the bounds obtained from the CDH method depends on the choice of
a certain function, usually called background field. The optimal background
field yielding the lowest upper bounds has again to be determined
numerically. A direct computation from the Euler-Lagrange equations (as for
instance in ref. \onlinecite{Ierley06} for infinite Prandtl number convection) is
cumbersome, which has prompted several authors \cite{Wen13,Wen15,Fantuz16} to
search for alternatives. This paper presents yet another alternative. It employs
the same constraints on the turbulent flow derived from the Navier-Stokes
equation as previous methods but uses a reformulation which allows to
numerically find the optimal background field through the solution of a
semidefinite program (SDP) which is simpler to implement than the SDP directly
derived from the CDH method solved in ref. \onlinecite{Fantuz16}.

Apart from these technical issues, the main purpose of this paper is to explore
the possibility of obtaining interesting bounds on energies. Previous work on
bounds focused on dissipation or transport, but other properties are of interest
as well. A bound on dissipation implies a bound on energy set by the maximal
amount of energy which can exist in the least dissipative flow mode without
violating the bound on dissipation. This argument is only based on the
definitions of energy and dissipation and does not take into account that the
flow field has to obey the Navier-Stokes equation. It is therefore of interest
to see whether the constraints used in deriving bounds on dissipation can lead
to improved bounds on energies. Since the expression for energy contains fewer
derivatives and is less singular than the dissipation, one can hope that it is
in some way simpler to find tight bounds on energies.

Section \ref{problem} starts with a precise statement of the problem. The
general idea of the employed method is worked out in section \ref{infinite_Pr}
taking as example upper bounds for the heat transport in infinite Prandtl number
convection. The results are extended to general Prandtl number in section
\ref{general_Pr}. Section \ref{energy} finally presents bounds on the poloidal
energy.

\section{Statement of the problem}
\label{problem}

Consider a plane layer of thickness $h$ with bounding planes perpendicular to
the $z-$axis, infinitely extended in the $x-$ and $y-$directions, and with
gravitational acceleration acting in the direction of negative $z$. The layer is
filled with fluid of density $\rho$, kinematic viscosity $\nu$, thermal
diffusivity $\kappa$, and thermal expansion coefficient $\alpha$. Top and bottom
boundaries are held at the fixed temperatures $T_\mathrm{top}$ and
$T_\mathrm{top} + \Delta T$, respectively. We will consider the equations of
evolution immediately in nondimensional form, choosing for units of length,
time, and temperature deviation from $T_\mathrm{top}$ the quantities $h$,
$h^2/\kappa$ and $\Delta T$. With this choice, the equations within the
Boussinesq approximation for the fields of velocity $\bm v(\bm r,t)$,
temperature $T(\bm r,t)$ and pressure $p(\bm r,t)$ become:

\begin{eqnarray}
\frac{1}{\mathrm{Pr}} \left( \partial_t\bm v + \bm v \cdot \nabla \bm v \right) &=& 
-\nabla p + \mathrm{Ra} \theta \bm{\hat z} + \nabla^2 \bm v
\label{eq:NS} \\
\partial_t \theta + \bm v \cdot \nabla \theta -v_z &=& \nabla^2 \theta
\label{eq:Temp} \\
\nabla \cdot \bm v &=& 0
\label{eq:conti} 
\end{eqnarray}

In these equations, $T=\theta+1-z$, so that $\theta$ represents the deviation
from the conduction profile. The Prandtl number $\mathrm{Pr}$ and the
Rayleigh number $\mathrm{Ra}$ are given by
\begin{equation}
\mathrm{Pr}=\frac{\nu}{\kappa} ~~~,~~~ 
\mathrm{Ra}=\frac{g \alpha \Delta T h^3}{\kappa \nu}
\end{equation}
and $\bm{\hat z}$ denotes the unit vector in $z-$direction.
The conditions at the boundaries $z=0$ and $z=1$ on the temperature are that 
$\theta=0$. Both stress free and no slip conditions will be investigated. At
stress free boundaries,
$\partial_z v_x = \partial_z v_y = v_z = 0$, whereas
$\bm v =0$ on no slip boundaries.

It is convenient to decompose $\bm v$ into poloidal and toroidal scalars $\phi$
and $\psi$ so that
$\bm v = \nabla \times \nabla \times (\phi \bm{\hat z}) + \nabla \times (\psi \bm{\hat z})$
which automatically fulfills $\nabla \cdot \bm v = 0$. The $z-$component of
the curl and the $z-$component of the curl of the curl of eq. (\ref{eq:NS})
yield the equations of evolution for $\phi$ and $\psi$,

\begin{eqnarray}
\frac{1}{\mathrm{Pr}} \left( \partial_t \nabla^2 \Delta_2 \phi
+ \bm{\hat z} \cdot \nabla \times \nabla \times \left[ (\nabla \times \bm v) \times \bm v \right] \right)
&=&
\nabla^2 \nabla^2 \Delta_2 \phi - \mathrm{Ra} \Delta_2 \theta 
\label{eq:phi}  \\
\frac{1}{\mathrm{Pr}} \left( \partial_t  \Delta_2 \psi
- \bm{\hat z} \cdot \nabla \times \left[ (\nabla \times \bm v) \times \bm v \right] \right)
&=&
\nabla^2 \Delta_2 \psi
\label{eq:psi} 
\end{eqnarray}

with $\Delta_2 = \partial^2_x + \partial^2_y$. For brevity, $\bm v$ is not
replaced by its expression in terms of $\phi$ and $\psi$ in these equations. The
stress free boundary conditions translate into 
$\phi = \partial^2_z \phi = \partial_z \psi = 0$ 
and the no slip boundary conditions become
$\phi = \partial_z \phi = \psi = 0$.

In the following, several types of averages will be important: the average over
the entire volume, denoted by angular brackets without subscript,
the average over an arbitrary
plane $z=\mathrm{const.}$, denoted by $\langle ... \rangle _A$, the average 
over a particular plane $z=z_0$, denoted by $\langle ... \rangle _{A,z=z_0}$,
and the average over time, which will be signaled by an overline.

The dot product of $\bm v$ with eq. (\ref{eq:NS}), followed by a volume average, leads to
\begin{equation}
\partial_t \langle \frac{1}{2} \bm v^2 \rangle =
- \mathrm{Pr} \mathrm{Ra} \langle \theta \Delta_2 \phi \rangle
- \mathrm{Pr} \langle |(\bm{\hat z}\times \nabla) \nabla^2 \phi|^2 
             + |\nabla \partial_x \psi|^2 + |\nabla \partial_y \psi|^2 \rangle
.
\label{eq:v2}
\end{equation}

Note that $v_z = - \Delta_2 \phi$. Multiplication of eq. (\ref{eq:Temp}) with
$\theta$ and a subsequent volume average yields
\begin{equation}
\partial_t \langle \frac{1}{2} \theta^2 \rangle = 
- \langle \theta \Delta_2 \phi \rangle - \langle |\nabla \theta|^2 \rangle
.
\label{eq:theta2}
\end{equation}
The average of eq. (\ref{eq:Temp}) over planes is noted for later reference:
\begin{equation}
\partial_t \langle \theta \rangle_A =
\langle \partial_z(\theta \Delta_2 \phi)\rangle_A + \langle \partial^2_z \theta \rangle_A
.
\label{eq:theta3}
\end{equation}

We will seek below bounds on the Nusselt number $\mathrm{Nu}$, defined as the
time averaged heat transport divided by the heat conducted if the fluid is at
rest. $\mathrm{Nu}-1$ can be determined from the vertical derivative of $\theta$
averaged over top or bottom boundaries:
$\mathrm{Nu}-1 = - \langle \partial_z \overline {\theta} \rangle_{A,z=0}
= - \langle \partial_z \overline {\theta} \rangle_{A,z=1}$.
Integration of eq. (\ref{eq:theta3}) from $z=0$ to any finite $z$, followed by
time averaging, shows that 
$\langle \overline {\theta \Delta_2 \phi} \rangle_A + \langle \partial_z
\overline {\theta} \rangle_A$
is identical at all height $z$, and in particular equal to
$\langle \partial_z \overline {\theta} \rangle_{A,z=0}$
because $\Delta_2 \phi = 0$ at $z=0$. An integration of
$\langle \overline {\theta \Delta_2 \phi} \rangle_A + \langle \partial_z
\overline {\theta} \rangle_A  = 
\langle \partial_z \overline {\theta} \rangle_{A,z=0}$
over all $z$ yields
$\langle \overline {\theta \Delta_2 \phi} \rangle =
\langle \partial_z \overline {\theta} \rangle_{A,z=0} =
- \mathrm{Nu} + 1$,
which yields an alternative expression for $\mathrm{Nu}-1$, and by virtue of
eq. (\ref{eq:theta2}) one also has
$\mathrm{Nu}-1 = \langle \overline{|\nabla \theta|^2} \rangle$

The second quantity for which we will seek bounds is the energy in the poloidal
part of the velocity field, $E_\mathrm{pol}$, given by
\begin{equation}
E_\mathrm{pol} = 
\langle \frac{1}{2} |\overline{\nabla \times \nabla \times (\phi \bm{\hat z})}|^2 \rangle
=\langle \frac{1}{2} |\overline{(\bm{\hat z} \times \nabla) \times \nabla \phi}|^2 \rangle
.
\label{eq:Epol}
\end{equation}
Because of the volume average, $E_\mathrm{pol}$ is strictly speaking an energy
density.

Considerable simplification results in the limit of infinite Prandtl number. The
time derivative and the advection term disappear from eq. (\ref{eq:NS}) in this
limit. Eq. (\ref{eq:psi}) then implies that $\psi=0$ and the momentum equation
becomes a diagnostic equation for $\phi$ alone:
\begin{equation}
\nabla^2 \nabla^2 \Delta_2 \phi - \mathrm{Ra} \Delta_2 \theta = 0
.
\label{eq:theta-phi}
\end{equation}
The temperature equation and the expressions for $\mathrm{Nu}$ and
$E_\mathrm{pol}$ as well as the boundary conditions are not affected by setting
$\mathrm{Pr}$ to infinity.

\section{Bounds for convection at infinite Prandtl number}
\label{infinite_Pr}

\subsection{The method}
This section explains the essence of the bounding method. Convection at infinite
Prandtl number serves as a simple example for more general problems. The
objective function $Z$ defines the quantity for whose time average $\overline Z$
we wish to find a bound. For example, the choice 
$Z=\langle |\nabla \theta|^2 \rangle$
will yield bounds on $\overline Z = \mathrm{Nu}-1$.

For infinite $\mathrm{Pr}$, eq. (\ref{eq:theta-phi}) allows us to eliminate
$\phi$ in favor of $\theta$ and we are left with equations for $\theta$ only.
The objective function $Z$ also depends only on $\theta$. Let us choose test
functions $\varphi_n(z)$, $n=1...N$ which depend on $z$ only and project onto them
the temperature equation (\ref{eq:Temp}):
\begin{equation}
\partial_t \langle \varphi_n \theta \rangle = 
\langle \varphi_n \partial_z (\theta \Delta_2 \phi) \rangle
+ \langle \varphi_n \nabla^2 \theta \rangle
.
\label{eq:projection}
\end{equation}

We now build a functional $G(\lambda_1 ... \lambda_N,\lambda_R,\theta)$ as a
linear combination of the right hand sides of the equations
(\ref{eq:projection}) and (\ref{eq:theta2}):
\begin{equation}
G(\lambda_1 ... \lambda_N,\lambda_R,\theta)=
\sum_{n=1}^N \lambda_n \left[ \langle \varphi_n \partial_z (\theta \Delta_2 \phi) \rangle
+ \langle \varphi_n \nabla^2 \theta \rangle \right]
+ \lambda_R \left[
\langle \theta \Delta_2 \phi \rangle + \langle |\nabla \theta|^2 \rangle
\right]
.
\end{equation}
The variable $\phi$ has been left in the expression for $G$ for better readability but it
does not appear as an argument of $G$ because of eq. (\ref{eq:theta-phi}) and
$\Delta_2 \phi = \mathrm{Ra} (\nabla^2 \nabla^2)^{-1} \Delta_2 \theta$,
where $(\nabla^2 \nabla^2)^{-1}$ is the inverse bi-Laplacian equipped with the
boundary conditions for $\phi$.

Assume that we know a set of coefficients $\lambda_1 ...
\lambda_N,\lambda_R$ and an additional number $\lambda_0$ such that for our
chosen objective function $Z(\theta)$ and all fields $\theta(\bm r)$ obeying the
boundary conditions for $\theta$, the following inequality holds:
\begin{equation}
-Z+\lambda_0+G(\lambda_1 ... \lambda_N,\lambda_R,\theta) \geq 0
.
\label{eq:p7Mitte}
\end{equation}
If this inequality is satisfied for every admissible $\theta$, it is in
particular satisfied by a solution of the equation of evolution. For such a
solution and a set of $\lambda$'s as specified above, we then have
\begin{eqnarray}
0 &=& G - \sum_{n=1}^N \lambda_n \partial_t \langle \varphi_n \theta \rangle
+ \lambda_R \partial_t \langle \frac{1}{2} \theta^2 \rangle 
\nonumber \\
 &\geq& Z - \lambda_0 + \partial_t \left[
-\sum_{n=1}^N \langle \varphi_n \theta \rangle + \lambda_R \langle \frac{1}{2} \theta^2 \rangle
\right]
.
\end{eqnarray}
Taking
the time average of the expression to the right of the inequality sign, we find
$\overline Z \leq \lambda_0$.
We see that we obtain an upper bound for $\overline Z$ as soon as we know a set
of coefficients $\lambda$ so that eq. (\ref{eq:p7Mitte}) is satisfied. The terms in $G$
are at most quadratic in $\theta$. If we wish to find upper bounds for
dissipation or energy, $Z$ is quadratic in $\theta$, too. If $\theta$ is
discretized, say, by spectral decomposition, the left hand side of inequality
(\ref{eq:p7Mitte}) is a quadratic form in the expansion coefficients of
$\theta$. Finding the sharpest bound on $\overline Z$ then amounts to minimizing
$\lambda_0$ subject to the condition that the matrix appearing in the
discretized form of inequality (\ref{eq:p7Mitte}) be positive semidefinite. This
problem has precisely the form of a semidefinite program.

In order to arrive at an implementable formulation, we have to select a
discretization for $\theta$ and test functions $\varphi_n$. A decomposition into
$N$ Chebychev polynomials $T_n$ for the $z-$ direction and into plane waves
in $x$ and $y$ is chosen as a
representation of $\theta$:
\begin{equation}
\theta = \sum_{n=1}^N \sum_{k_x} \sum_{k_y} {\hat \theta}_{n,k_x,k_y}
T_n(2z-1) e^{i(k_x x + k_y y)}
.
\label{eq:expansion}
\end{equation}
The summation limits for $k_x$ and $k_y$ are not specified because they will
turn out to be irrelevant. The coefficients ${\hat \theta}_{n,k_x,k_y}$ for any
given $k_x,k_y$ are not independent of each other because of the boundary
conditions on $\theta$ at $z=0$ and $z=1$. Only $N_f$ of the $N$ coefficients
are free. In order to enforce $\theta=0$ at $z=0$ and $z=1$, it is enough to set
$N=N_f+2$. However, as reported below, it was found numerically advantageous to
also explicitly enforce $\partial^2_z \theta = 0$ at $z=0$ and $z=1$ (a relation which
follows from eq. (\ref{eq:Temp}) and all boundary conditions considered here), in which
case $N=N_f+4$.

The test functions $\varphi_n$ are chosen as delta functions centered at the collocation
points in common use in time integration methods based on Chebychev
discretization:
\begin{eqnarray}
z_n &=& \frac{1}{2} \left[ 1+\cos \left( \pi \frac{n-1}{N-1} \right) \right]
~~~,~~~ n=1 ... N   \label{eq:colloc}\\
\varphi_n(z) &=& \delta(z-z_n)
.
\end{eqnarray}

Inserting all this into eq. (\ref{eq:p7Mitte}) leads to an inequality of the
form
\begin{equation}
-\sum_{k_x,k_y} \bm x^*_{k_x,k_y} \bm Z_k \bm x_{k_x,k_y} + \lambda_0
+\sum_{n=1}^N \lambda_n \sum_{k_x,k_y} \bm x^*_{k_x,k_y} \bm G_{nk} \bm x_{k_x,k_y}
+\bm x^*_{k_x,k_y} \lambda_R \bm E_k \bm x_{k_x,k_y}
\geq 0
.
\label{eq:p7discret}
\end{equation}
The superscripted star denotes complex conjugation. The matrices $\bm Z_k$, $\bm
E_k$ and
$\bm G_{nk}$ are real. Their entries depend on $k_x$ and $k_y$ only in the
combination $k^2=k_x^2+k_y^2$ and there are no terms coupling different $k$.
Furthermore, these matrices are symmetric. The vectors 
$\bm x_{k_x,k_y}$ are constructed as
$\bm x_{k_x,k_y} = \left( 1, {\hat x}_{1,k_x,k_y},{\hat x}_{2,k_x,k_y},
... , {\hat x}_{N_f,k_x,k_y} \right)$
where the ${\hat x}_{n,k_x,k_y}$ are linear combinations of the 
${\hat \theta}_{n,k_x,k_y}$ in eq. (\ref{eq:expansion}). They appear in an
expansion analogous to eq. (\ref{eq:expansion}) as the coefficients of $N_f$
linear combinations of the first $N$ Chebychev polynomials satisfying the
boundary conditions.
The 1 in the first component of $\bm x_{k_x,k_y}$
is necessary in order to accommodate the term linear
in $\theta$ in the $k=0$ summand in eq. (\ref{eq:p7discret}).
It also allows us to write $\lambda_0$ as
$\sum_{k_x,k_y} \bm x^*_{k_x,k_y} \lambda_0 \bm H_{k} \bm x_{k_x,k_y}$
with $\bm H_k = 0$ for $k \neq 0$ and all entries of $\bm H_0$ are zero except
that $H_{0,1,1}=1$.

Inequality (\ref{eq:p7discret}) therefore has the form
\begin{equation}
\sum_{k_x,k_y} \bm x^*_{k_x,k_y}
\left[ -\bm Z_k + \lambda_0 \bm H_k + \sum_{n=1}^N \lambda_n \bm G_{nk} + \lambda_R \bm E_k \right]
\bm x_{k_x,k_y} \geq 0
.
\end{equation}
For $\lambda_0$ to serve as an upper bound, this inequality must be satisfied
for all $\bm x_{k_x,k_y}$, which requires that 
$\left[ -\bm Z_k + \lambda_0 \bm H_k + \sum_{n=1}^N \lambda_n \bm G_{nk} + \lambda_R \bm
E_k \right]$
be positive semidefinite for all $k$. The SDP to be solved is in summary:
\begin{equation}
\begin{aligned}
\text{minimize } & \lambda_0  \\
\text{subject to } & 
-\bm Z_k + \lambda_0 \bm H_k + \sum_{n=1}^N \lambda_n \bm G_{nk} + \lambda_R \bm E_k 
\succeq 0 ~~~~~~ \text{for all }k
\end{aligned}
\label{eq:SDP}
\end{equation}
where $\bm M \succeq 0$ means that the matrix $\bm M$ is positive semidefinite.

The symmetry about the midplane can be exploited to alleviate the numerical task
of solving the SDP. This symmetry suggests to construct linear combinations of
the constraints (\ref{eq:projection}) in the form of equations for
$\partial_t \left( \langle \varphi_i \theta \rangle \pm 
\langle \varphi_{N+1-i} \theta \rangle \right) 
~,~ i=1 ... N/2 $.
It turns out that only the equations with the minus sign constrain $\lambda_0$,
the other equations are redundant. Symmetry affects also the parity of the most
dangerous functions which most readily violate the inequality constraint of the
SDP. Expanding $\theta$ in the functions 
$T_{n+2}(2z-1) - T_n(2z-1)$, which are by construction zero at $z=0$ and $z=1$,
one only needs to retain odd $n$ for $k=0$ and even $n$ for $k \neq 0$. The
other combinations do not affect the best bound.

It may seem like a formidable task to enforce the inequality condition in
(\ref{eq:SDP}) for all $k$. In fact, it is not. Barring any special symmetry
or degeneracy in the matrices involved,
the inequality will hold as a strict inequality for almost all $k$ and as an equality only
for a finite set of $k$, the active set in the parlance of optimization. The active
set tends to be small. As an example, pretend that we are looking for a bound on
dissipation so that $Z=\langle |\nabla \theta|^2 \rangle$. We should be able to
prove that dissipation is zero for $\mathrm{Ra}$ below onset. This requires
$\lambda_0=0$. We can find a suitable set of coefficients by setting
$\lambda_1=\lambda_2=...=\lambda_N=0$ and only $\lambda_R \neq 0$. 
Eq. (\ref{eq:p7Mitte}) becomes
$
\lambda_R \langle \theta \Delta_2 \phi \rangle + 
(\lambda_R-1) \langle |\nabla \theta|^2 \rangle \geq 0
$.
The last term suggests that we are best off choosing $\lambda_R$ positive and
large compared with 1, so that the condition
\begin{equation}
\langle \theta \Delta_2 \phi \rangle + \langle |\nabla \theta|^2 \rangle \geq 0
\label{eq:energy_stab}
\end{equation}
needs to be satisfied
for all admissible $\theta$ ($\phi$ being determined by eq.
(\ref{eq:theta-phi})). The left hand side of eq. (\ref{eq:energy_stab}) is equal
to $-\partial_t \langle \frac{1}{2} \theta^2 \rangle$, so that eq.
(\ref{eq:energy_stab}) is nothing but the condition for energy stability. The
$\theta$ which yields the largest possible left hand side is determined from a linear
Euler-Lagrange equation with coefficients independent of $k$, since the left
hand side itself has no $k$ dependent coefficients. The optimal $\theta$ has
therefore a single Fourier component. At the critical Rayleigh number, the
inequality condition in (\ref{eq:SDP}) is fulfilled with the equality sign for
one value of $k$ and as a strict inequality for all other $k$. At least at this
$\mathrm{Ra}$, the active set contains only one element. As $\mathrm{Ra}$ is
increased, the active set gradually becomes larger, adding one element after the
other.

All results presented in this paper were computed with an automated search for
the active set. As one is usually interested in some bound as a function of
$\mathrm{Ra}$, start with a small $\mathrm{Ra}$ and a tentative active set of
two elements: $k=0$, which is always necessary for any $\lambda_i \neq 0, 1 \leq
i \leq N$ because of the last term in eq. (\ref{eq:projection}), and an
additional $k \neq 0$. Then run an optimization over the nonzero $k$ which
maximizes $\lambda_0$, where $\lambda_0$ is the result of the SDP problem
\begin{equation}
\begin{aligned}
\text{minimize } & \lambda_0  \\
\text{subject to } & 
-\bm Z_k + \lambda_0 \bm H_k + \sum_{n=1}^N \lambda_n \bm G_k + \lambda_R \bm E_k 
\succeq 0 ~~~~~~ \text{for all $k$ in the active set}
\end{aligned}
\end{equation}
This maximization selects the wavenumbers $k$ which impose the most stringent
constraint on $\lambda_0$.
Once the maximum is found, it remains to check whether the active
set is complete. For this purpose, the matrix 
$\left[ -\bm Z_k + \lambda_0 \bm H_k + \sum_{n=1}^N \lambda_n \bm G_{nk} + \lambda_R \bm
E_k \right]$
is assembled for 1000 values of k regularly spaced on a logarithmic scale for $k$
between 0.1 an $10^4$, and for each matrix, a Cholesky decomposition is
attempted. If the matrix is positive semidefinite at all tested wavenumbers, the
calculation advances to the next higher $\mathrm{Ra}$ to compute the next bound. 
If the Cholesky decomposition fails, one has found a $k$ which violates the inequality
constraint in the SDP. The tentative active set is then enlarged by one element
and the maximization of $\lambda_0$ over the wavenumbers in the active set is
run anew. If the new maximal $\lambda_0$ is larger by at least $0.1\%$ than the
previous result, the new active set is accepted as such and $\mathrm{Ra}$ is
increased to compute the next bound. If the fractional increase is less than
$0.1\%$, the newly added $k$ is rejected as a false alarm and the old active set
is used as the starting point for the computation of the bound at the next
$\mathrm{Ra}$. This rejection is a useful procedure because tolerances in the
optimization together with round off errors accumulated during the Cholesky
decomposition may lead the method to spuriously signal a violation of the
constraint of positive semidefiniteness. The most straightforward way to
distinguish a spurious from a warranted detection of violation is to check
whether an enlarged set of wavenumbers leads to a more stringent constraint on
$\lambda_0$. If it does not, the enlargement was unnecessary.

The most time consuming step in this entire method is the solution of the SDP.
This was done with the Python interfaced version of the package cvxopt available
from cvxopt.org, using
either the internal solver of cvxopt or dsdp5. These codes implement central
path methods which have the reputation of being the best choice for small to
medium size problems. Resolutions of up to $N=512$ were used for the results
presented in this paper which are limited in
Rayleigh number not because of an overwhelming computational burden, but because
the SDP solvers failed to converge at large $\mathrm{Ra}$. The SDP package was
used as a black box which precluded a precise tracking of the problem, but
round off issues are the most likely culprit. Ierley et al. \cite{Ierley06}
report that they had to use a 96 digit floating point representation to cover
their interval of $\mathrm{Ra}$ in an implementation of the CDH method. The
package used for the present work did not allow one to vary floating point
types, so that the effect of the data type on convergence was not tested. The
internal solver of cvxopt allowed to treat slightly larger $\mathrm{Ra}$ than
dsdp5, and if both methods converged, they converged to the same result. 

Even though it is enough to enforce $\theta=0$ at the boundaries, it helped to
require $\partial_z^2 \theta = 0$ as well. Again, both sets of boundary
conditions lead to the same results if the SDP solver converges, but the
extended boundary conditions allow one to reach higher $\mathrm{Ra}$. The use of
Chebychev polynomials is not the root of the convergence problems. For free slip boundaries,
it is also convenient to choose 
$\varphi_n = \sin(n\pi z)$ and an expansion of $\theta$ in $\sin(n \pi z)$ instead
of Chebychev polynomials. This choice also leads to convergence problems at very
much the same $\mathrm{Ra}$.

Some relief is provided by an obvious modification of the method which consists
in integrating by parts in eq. (\ref{eq:projection}) in order to obtain

\begin{equation}
\partial_t \langle \varphi_n \theta \rangle =
-\langle (\partial_z \varphi_n) \theta \Delta_2 \phi \rangle
- \langle (\partial_z \varphi_n) (\partial_z \theta) \rangle
+\varphi_n(1) \langle \partial_z \theta \rangle_{A,z=1}
-\varphi_n(0) \langle \partial_z \theta \rangle_{A,z=0}
\label{eq:Abl_projection}
\end{equation}
and to choose the test functions $\varphi_n$ such that 
$\partial_z \varphi_n = \delta(z-z_n)$ with the Chebychev collocation points
defined in eq. (\ref{eq:colloc}). In order to preserve the symmetry about
$z=1/2$, one can express the test function at the boundaries as
\begin{equation}
\begin{aligned}
\varphi(0) &=& \varphi(\frac{1}{2}) + \int_\frac{1}{2}^0 \partial_z \varphi dz
&=& \varphi(\frac{1}{2}) - \sum_{n=1}^{N/2} \lambda_n \\
\varphi(1) &=& \varphi(\frac{1}{2}) + \int_\frac{1}{2}^1 \partial_z \varphi dz
&=& \varphi(\frac{1}{2}) + \sum_{n=N/2+1}^N \lambda_n 
\end{aligned}
\end{equation}
for $\varphi(z) = \sum_{n=1}^N \lambda_n \varphi_n(z)$ with $N$ even. We can
therefore replace the functional $G$ with $G'$ defined as
\begin{equation}
\begin{aligned}
G'(\lambda_1 ... \lambda_N,\lambda_R,\theta) =
&
-\sum_{n=1}^N \lambda_n \langle \delta(z-z_n) \left[
\theta \Delta_2 \phi + \partial_z \theta \right] \rangle
+\varphi(\frac{1}{2}) \left( \langle \partial_z \theta \rangle_{A,z=1}
- \langle \partial_z \theta \rangle_{A,z=0} \right)
\\
 &
+\sum_{n=1}^{N/2} \lambda_n \langle \partial_z \theta \rangle_{A,z=0}
+\sum_{n=N/2+1}^N \lambda_n \langle \partial_z \theta \rangle_{A,z=1}
+ \lambda_R \left[
\langle \theta \Delta_2 \phi \rangle + \langle |\nabla \theta|^2 \rangle
\right]
\end{aligned}
\end{equation}
and we must obtain identical results if we replace $G$ with $G'$ in eq.
(\ref{eq:p7Mitte}). Because of the symmetry of the optimal test function,
$\varphi(z)=-\varphi(1-z)$ so that $\varphi(\frac{1}{2})=0$, and the second term
in $G'$ disappears. $G'$ still looks more complicated than $G$ but does
not contain second derivatives. As a result, the implementation using $G'$ has
better convergence properties and is preferred for this reason.

Finally, rescaling $\lambda_0$ improves convergence. If the optimal $\lambda_0$
approximately equals $c \mathrm{Ra}^\gamma$ with some constants $c$ and $\gamma$,
it helps to use the variable $\lambda'_0=\lambda_0/(c \mathrm{Ra}^\gamma)$
instead of $\lambda_0$ itself as an argument of the functional $G'$.

The central path methods also solve the dual
problem. This part of the output allows one to compute the derivative of the
optimal $\lambda_0$
with respect to problem parameters such as the
wavenumbers $k$ (see example 5.13 in ref. \onlinecite{Boyd04}). This opens the possibility
of using a gradient based method for maximizing $\lambda_0$ over $k$. This was
done with a Newton-BFGS method. However, in order to run advantageously, this
method needs derivatives with a higher precision than what the SDP solver
sometimes provides. If the progress of Newton-BFGS was too slow, it
was replaced by the gradient free Nelder-Mead optimization.

\subsection{Relation with other methods}

The background field or CDH method can be directly mapped onto the method of
this section. It proceeds by using a different variable for the temperature
variation, defined by
$T(\bm r,t) = \tau(z) + \theta'(\bm r,t)$
where the background field $\tau$ is chosen such that $\tau(0)=1$, $\tau(1)=0$,
and $\theta'=0$ on the boundaries. It follows from the temperature equation
(\ref{eq:Temp}) that
\begin{equation}
\partial_t \langle \frac{1}{2} \theta'^2 \rangle = 
- \langle \partial_z \theta' \partial_z \tau \rangle 
- \langle v_z \theta' \partial_z \tau \rangle
- \langle |\nabla \theta'|^2 \rangle
.
\label{eq:theta'2}
\end{equation}
The CDH method now consists in finding $\tau(z)$, $\lambda_0$ and $\lambda_R$
such that
\begin{equation}
-Z + \lambda_0 - \lambda_R \left(
\langle \partial_z \theta' \partial_z \tau \rangle
+\langle v_z \theta' \partial_z \tau \rangle
+\langle |\nabla \theta'|^2 \rangle
\right) \geq 0
\label{eq:CDH_ineq}
\end{equation}
for all admissible $\theta'$ and $v_z$, in which case 
$\overline Z \leq \lambda_0$. This problem is not directly an SDP if $\tau$
appears quadratically in $Z$. It is then necessary to form a Schur
complement in order to recover an optimization problem in SDP form
\cite{Fantuz16}. This complicates a direct implementation of the CDH method
compared with the method presented above. However, the bounds obtained from both
methods must be identical, because after the identification
$\varphi(z) = \sum \lambda_n \varphi_n(z) = \lambda_R  (\tau-1+z)$
one recognizes equations (\ref{eq:CDH_ineq}) and (\ref{eq:p7Mitte}) to be the
same. This also means that the boundary layer structure typical of optimal
background fields will also appear in $\varphi(z)$, and we also see that
$\varphi$ obeys $\varphi(0)=\varphi(1)=0$.

As a further variation of the bounding problem, and in order to make contact
with ref. \onlinecite{Seis15}, consider the test function
$\varphi_S$ with $\varphi_S=1$ for
$z \leq l$ and $\varphi_S=0$ for $l < z \leq 1$ for some $l$ with $0 < l < 1$.
We then obtain
\begin{equation}
\partial_t \langle \varphi_S \theta \rangle = 
-\int_0^l \langle \theta \Delta_2 \phi + \partial_z \theta \rangle_A dz
-\langle \partial_z \theta \rangle_{A,z=0}
\end{equation}
whose discretization is a suitable linear combination of eqs.
(\ref{eq:Abl_projection}).
A bound for any time averaged $Z$ in the form 
$\overline Z \leq \lambda_0$ is then obtained if one finds $\lambda_0$, $l$ and
$\lambda_R$ such that
\begin{equation}
-Z + \lambda_0 -\int_0^l \langle \theta \Delta_2 \phi + \partial_z \theta \rangle_A dz
-\langle \partial_z \theta \rangle_{A,z=0}
+\lambda_R \left[
\langle \theta \Delta_2 \phi \rangle + \langle |\nabla \theta|^2 \rangle
\right] \geq 0
\end{equation}
for all admissible $\theta$ and $\phi$. Seis \cite{Seis15} further restricts the
problem to $\lambda_R=1$ and shows how to solve it for general Prandtl number.

The functional $G$ contains a linear combination of expressions which are the
right hand sides of the time evolution equations of terms which are linear or
quadratic in the flow variables $\theta$, $\phi$, $\psi$. One could add linear
combinations of the right hand sides of time evolution equations for higher
moments, which would be equivalent to the auxiliary function approach
\cite{Tobasc18,Golusk18}. This leads to an optimization problem with a
constraint of the form that some polynomial must always be positive. An SDP can
then only be formulated after relaxing this constraint to the less restrictive
requirement that this polynomial is a sum of squares \cite{Tobasc18,Golusk18}.

\subsection{Heat transport}

\begin{figure}
\includegraphics[width=8cm]{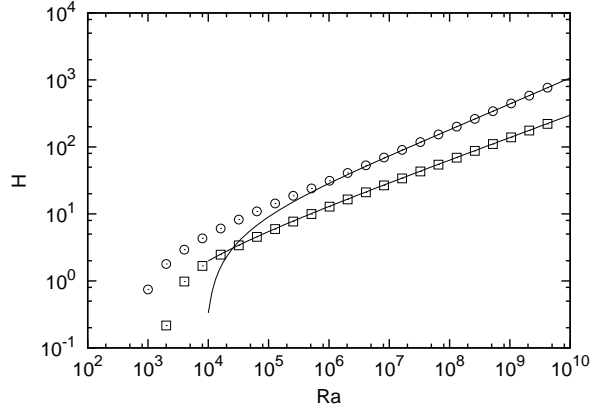}
\caption{
The optimal upper bound $H$ for the advective heat transport $\mathrm{Nu}-1$ as
a function of $\mathrm{Ra}$ for stress free (circles) and no slip (squares)
boundary conditions and infinite $\mathrm{Pr}$.
The continuous lines show formulae obtained from fits in ref. \onlinecite{Ierley06},
which are $H=0.101 \times \mathrm{Ra}^{0.4} + 0.70965 \times \mathrm{Ra}^{0.2} -
7.166-1$ for free slip boundaries and
$H=0.139 \times \mathrm{Ra}^{1/3}-1$ for no slip boundaries.}
\label{fig_Nu_bounds_inf}
\end{figure}

\begin{figure}
\includegraphics[width=8cm]{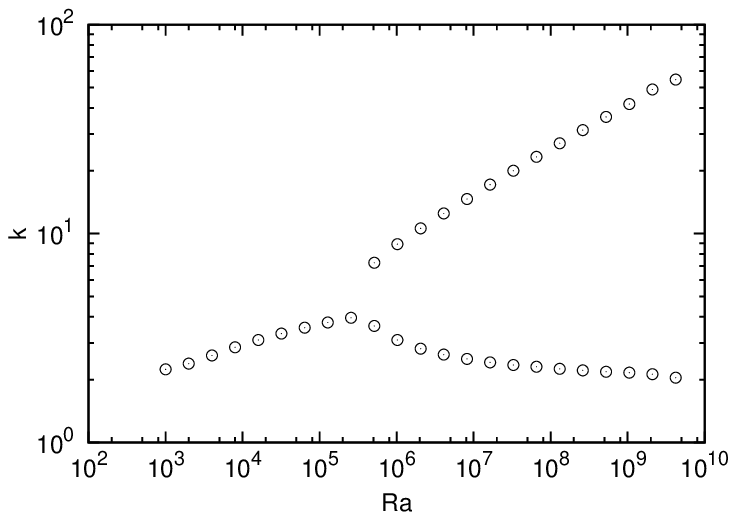}
\caption{
Wavenumbers $k$ in the active set for the optimization of the bound $H$ for free
slip boundary conditions and infinite $\mathrm{Pr}$.}
\label{fig_Nu_bifurc_fs}
\end{figure}

We will now consider bounds on the heat transport as a validation of the method
presented in this section. These bounds are obtained by setting 
$Z=\langle |\nabla \theta|^2 \rangle$, or
$Z=\langle -\theta \Delta_2 \phi \rangle$ or
$Z=\langle - \partial_z \theta \rangle_{A,z=0}$.
With all these three choices, $\overline Z = \mathrm{Nu}-1$, and the results
obtained with the method described above are indeed the same. Fig.
\ref{fig_Nu_bounds_inf} shows the optimal bound $H$ on $\mathrm{Nu}-1$ found for
both stress free and no slip boundaries. The distinction between the two
boundary conditions appears only in the boundary conditions for the inverse
bi-Laplacian in eq. (\ref{eq:theta-phi}). 

For both boundary conditions, Ierley et al. \cite{Ierley06} have obtained bounds 
from numerical optimization within the CDH method. Approximate expressions for 
those bounds are also included in fig. \ref{fig_Nu_bounds_inf}. These bounds
conform with the analytical result that $H \leq c_f \mathrm{Ra}^{5/12}$ for free
slip boundaries \cite{Whiteh12} with some constant $c_f$. Lower bounds
are obtained if the number of elements in the active set is limited. The formula
quoted in the caption of fig. \ref{fig_Nu_bounds_inf} is valid for up to two
wavenumbers in the active set, which is appropriate for the range of
$\mathrm{Ra}$ in fig. \ref{fig_Nu_bounds_inf}, and which is a more accurate fit
than the universally valid bound proportional to $\mathrm{Ra}^{5/12}$. 
For no slip boundaries, it is known \cite{Nobili17} that the CDH method cannot
yield bounds better than proportional to 
$\mathrm{Ra}^{1/3} (\ln \mathrm{Ra})^{1/15}$. This agrees with the result of ref.
\onlinecite{Ierley06} apart from the virtually undetectable factor
$(\ln \mathrm{Ra})^{1/15}$.

Bounds obtained from the MHB method can
be found in ref. \onlinecite{Vitano98} and in tabulated form in Vitanov's doctoral
thesis \cite{Vitano98b}. The numbers in the thesis match the results shown in
fig. \ref{fig_Nu_bounds_inf} to four digits accuracy.

Fig. \ref{fig_Nu_bifurc_fs} shows the wavenumbers included in the active set for
the example of free slip boundaries. These are not exactly the same as those
given in the analogous figure 12b of ref. \onlinecite{Ierley06} which shows an
additional wavenumber at $\mathrm{Ra} > 10^9$. This difference simply comes from
the tolerances set in the automated search of the wavenumbers. The additional
wavenumber for $\mathrm{Ra}$ above $10^9$ does increase the bound $H$, but by
less than $0.1\%$ which is why it was rejected and does not appear in fig.
\ref{fig_Nu_bifurc_fs}.

\section{Bounds on heat transport at general Prandtl number}
\label{general_Pr}

\begin{figure}
\includegraphics[width=8cm]{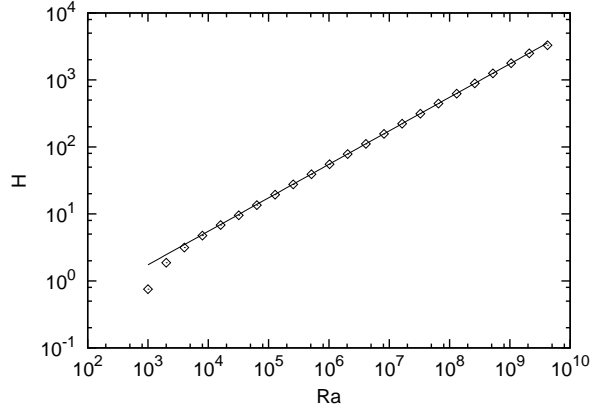}
\caption{
The optimal upper bound $H$ for the advective heat transport $\mathrm{Nu}-1$ as
a function of $\mathrm{Ra}$ for stress free boundaries and general $\mathrm{Pr}$. The 
continuous line indicates the power law $0.055 \times \mathrm{Ra}^{1/2}$.}
\label{fig_Nu_bounds_general}
\end{figure}

\begin{figure}
\includegraphics[width=8cm]{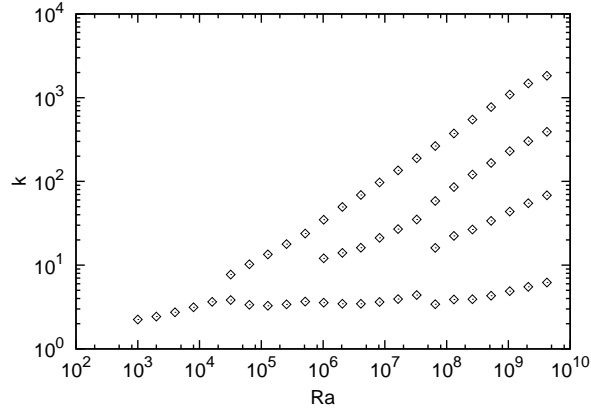}
\caption{
Wavenumbers $k$ in the active set for the optimization of the bound $H$ for free
slip boundary conditions and general $\mathrm{Pr}$.}
\label{fig_Nu_bifurc_general}
\end{figure}

For finite $\mathrm{Pr}$, one has to return to the general equations
(\ref{eq:phi}) and (\ref{eq:psi}). The equations of evolution for 
$\langle \theta \rangle_A$ and $\langle \frac{1}{2} \theta^2 \rangle$,
equations (\ref{eq:theta3}) and (\ref{eq:theta2}) remain unchanged, but the dot
product of the momentum equation (\ref{eq:NS}) with $\bm v$, averaged over the
entire volume, now yields the additional non trivial information in eq.
(\ref{eq:v2}) that
\begin{equation}
\partial_t \langle \frac{1}{2} \bm v^2 \rangle =
-\mathrm{Pr} \mathrm{Ra} \langle \theta \Delta_2 \Phi \rangle
-\mathrm{Pr} \langle |(\bm{\hat z} \times \nabla) \nabla^2 \phi |^2 \rangle
-\mathrm{Pr} D_\mathrm{np}
\label{eq:v2mitD}
\end{equation}
with the toroidal dissipation
\begin{equation}
D_\mathrm{np} = \langle |\nabla \partial_x \psi|^2 + |\nabla \partial_y \psi|^2
\rangle.
\end{equation}
The subscript stands for non poloidal, because if periodic boundary conditions
are imposed on $\psi$, this term has to include not only the
dissipation due to the toroidal field 
but also the dissipation due to a horizontal mean flow \cite{Schmit92}.
Since $D_\mathrm{np} \geq 0$, it will be convenient to introduce a new variable
$d$ defined by $d^2 = D_\mathrm{np}$.

We can now follow the same line of thought as in the previous section and
construct a functional $F$ as
\begin{equation}
\begin{aligned}
F(\lambda_1 ... \lambda_N,\lambda_R,\lambda_E,\theta,\phi,d) = 
&
\sum_{n=1}^N \lambda_n \langle \varphi_n \left[ \partial_z (\theta \Delta_2 \phi)
+ \partial_z^2 \theta \right] \rangle
+\lambda_R \left[
\langle \theta \Delta_2 \phi \rangle + \langle |\nabla \theta|^2 \rangle
\right]
\\
 &
+\lambda_E \left[ \mathrm{Ra} \langle \theta \Delta_2 \phi \rangle
+\langle |(\bm{\hat z} \times \nabla) \nabla^2 \phi |^2 \rangle
\right]
+\lambda_E d^2
\end{aligned}
\end{equation}
and conclude that if we know $\lambda_0,\lambda_1 ...
\lambda_N,\lambda_R,\lambda_E$ such that
\begin{equation}
-Z + \lambda_0 + F(\lambda_1 ... \lambda_N,\lambda_R,\lambda_E,\theta,\phi,d) \geq 0
\label{eq:p7general}
\end{equation}
for all admissible $\theta$, $\phi$ and all $d$, we get the estimate
$\overline Z \leq \lambda_0$. The functional $F$ has to include both $\theta$
and $\phi$ as independent arguments because the relation (\ref{eq:theta-phi}) no
longer holds for finite $\mathrm{Pr}$. There is also an additional coefficient
$\lambda_E$ which has to be positive in order to satisfy the inequality
(\ref{eq:p7general}) for all $d$.

The bounding method was numerically implemented for stress free boundaries using
identical expansions for $\theta$ and $\phi$ into Chebychev polynomials and
Fourier series, and using the same test functions $\varphi_n$ as for infinite
$\mathrm{Pr}$ in the previous section. The optimizing fields $\theta$ and $\phi$
again have symmetry, which was exploited in the calculations with high
resolutions. When expanded in 
$T_{n+2}(2z-1)-T_n(2z-1)$, the Fourier components of both $\theta$ and $\phi$
contain only terms with even $n$ for $k>0$. There is no contribution to $\phi$
with $k=0$, and the zero wavenumber contribution to $\theta$ contains only terms
with odd $n$. Expressed differently, if
$\theta = \sum_{k_x,k_y} \tilde \theta_{k_x,k_y} e^{i(k_x x + k_y y)}$
and $k^2=k_x^2+k_y^2$, then
$\tilde \theta_{k_x,k_y} (z) = \tilde \theta_{k_x,k_y} (1-z)$
and
$\tilde \theta_{0,0} (z) = - \tilde \theta_{0,0} (1-z)$,
and likewise for the poloidal field. The optimal $\varphi(z)$ has the same
symmetry as for infinite $\mathrm{Pr}$.

Bounds $H$ for $\mathrm{Nu}-1$ are obtained by setting $Z$ to either of the
following expressions:
$\langle |\nabla \theta|^2 \rangle$, 
$\langle -\theta \Delta_2 \phi \rangle$, or
$\langle - \partial_z \theta \rangle_{A,z=0}$.
All three options yield identical results and they agree to four digits accuracy
with values listed in Vitanov's thesis \cite{Vitano98b}. The result is shown in
fig. \ref{fig_Nu_bounds_general}. The bounds obey approximately
$H=0.055 \times \mathrm{Ra}^{1/2}$. The active set now contains more
wavenumbers, as shown in fig. \ref{fig_Nu_bifurc_general}.

\section{Bounds on the velocity field}
\label{energy}

The previous sections investigated bounds on heat transport or thermal
dissipation, a quantity obtainable from the temperature field alone. The present
section deals with the velocity field, which is characterized first and foremost
by its energy. Rewriting eq. (\ref{eq:v2}), the dissipation of the velocity field is given by
\begin{equation}
\sum_{ij} \langle \overline{(\partial_j v_i)^2} \rangle = \mathrm{Ra} (\mathrm{Nu}-1)
\label{eq:E_budget}
\end{equation}
so that a bound on $\mathrm{Nu}-1$ also bounds the kinetic dissipation.
We can now invoke Poincar\'e's inequality, which states that 
$\int_0^1 (\partial_z f)^2 dz \geq \eta \int_0^1 f^2 dz$
for any function $f(z)$ defined on the interval $0 \leq z \leq 1$,
where $\eta$ is the smallest eigenvalue of the Laplace equation
$-\partial_z^2 g = \eta g$, if $g$ and $f$ obey the same
boundary conditions. For Dirichlet boundaries, $\eta = \pi^2$. Neumann boundary conditions 
do not uniquely select a solution. If $f$ obeys $\int_0^1 f(z)dz = 0$ in
addition to the Neumann conditions, one has again $\eta = \pi^2$. We now also
use the fact that the total dissipation is greater than the dissipation due to
the poloidal velocity field $\bm v_P$ alone, and that all components in
$\bm v_p = \partial_z \partial_x \phi \bm{\hat x}
+\partial_z \partial_y \phi \bm{\hat y}
-\Delta_2 \phi \bm{\hat z}$
either obey Dirichlet conditions in $z$, or satisfy Neumann conditions
and their integral over $z$ is zero:
\begin{eqnarray}
\sum_{ij} \langle \overline{(\partial_j v_i)^2} \rangle
& \geq & \sum_{ij} \langle \overline{(\partial_j v_{P,i})^2} \rangle
\nonumber \\
& \geq &
\langle \overline{(\partial_z v_{P,x})^2} + \overline{(\partial_z v_{P,y})^2}
+ \overline{(\partial_z v_{P,z})^2} \rangle
\nonumber \\
& \geq & \pi^2 \left( \langle \overline{v_{P,x}^2} \rangle + \langle
\overline{v_{P,y}^2} \rangle
+ \langle \overline{v_{P,z}^2} \rangle \right)
\end{eqnarray}
If $H$ is an
upper bound for $\mathrm{Nu}-1$, the poloidal energy is thus bounded as
\begin{equation}
\mathrm{Ra} H \geq 2 \pi^2 E_\mathrm{pol}
.
\label{eq:ineq_Ekin}
\end{equation}
For infinite $\mathrm{Pr}$, the velocity field is purely poloidal and poloidal
and total energies are identical.
For finite $\mathrm{Pr}$, the toroidal field and a possible mean flow also
contribute to the kinetic energy, but they do not appear in the constraints
employed in the formulation of the optimization problem in section
\ref{general_Pr}. We therefore cannot obtain any additional information about
the total energy, we can only hope to find tight bounds on the poloidal energy
with the technique described above. 
The bound (\ref{eq:ineq_Ekin})
derives solely from the energy budget (\ref{eq:E_budget}) and the definitions of
kinetic dissipation and energy. We expect to find tighter bounds by solving the
full optimization problem.

\begin{figure}
\includegraphics[width=8cm]{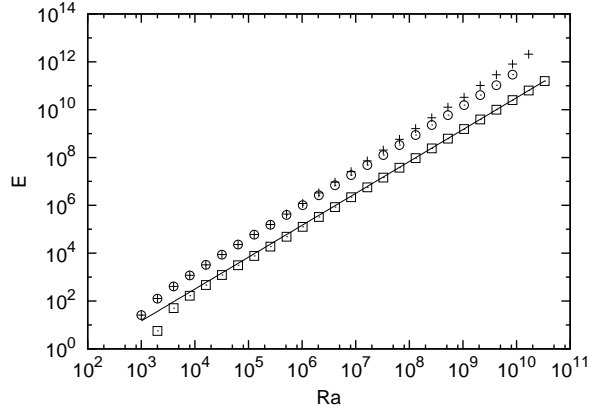}
\caption{
The optimal upper bound $E$ for the poloidal kinetic energy as
a function of $\mathrm{Ra}$ at infinite $\mathrm{Pr}$
for stress free (circles) and no slip (squares)
boundary conditions as well as for general $\mathrm{Pr}$ and
free slip boundaries (plus signs).
The continuous line indicates the power law $0.00147 \times \mathrm{Ra}^{4/3}$.}
\label{fig_Epol_bounds}
\end{figure}

\begin{figure}
\includegraphics[width=8cm]{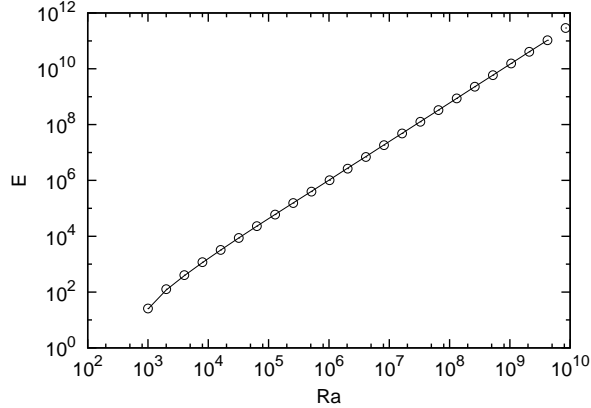}
\caption{
The optimal upper bound $E$ for the poloidal kinetic energy as
a function of $\mathrm{Ra}$ at infinite $\mathrm{Pr}$
and stress free boundaries.
The continuous line shows the function $0.033 \times \mathrm{Ra} H$, where $H$ is the
upper bound on the advective heat transport from fig. \ref{fig_Nu_bounds_inf}.}
\label{fig_Epol_fs}
\end{figure}

\begin{table}\centering
\begin{tabular}{ c | c | c | c }

 & $H$ & $E$ & $\mathrm{Ra} H/(2 \pi^2)$ \\

\hline

infinite $\mathrm{Pr}$, free slip & $ $ &
$0.033 \times \mathrm{Ra}H$ & $0.05066 \times \mathrm{Ra}H$ \\

\hline

infinite $\mathrm{Pr}$, no slip & $0.139 \times \mathrm{Ra}^{1/3}$ &
$0.00147 \times \mathrm{Ra}^{4/3}$ & $0.00704 \times \mathrm{Ra}^{4/3}$ \\

\hline

general $\mathrm{Pr}$, free slip & $0.055  \times \mathrm{Ra}^{1/2}$ &
$0.00106 \times \mathrm{Ra}^{3/2}$ & $0.002786 \times \mathrm{Ra}^{3/2}$ \\

\end{tabular}
\caption{Fitted dependencies on $\mathrm{Ra}$ of $H$ and $E$, the optimal upper
bounds on $\mathrm{Nu}-1$ and $E_\mathrm{pol}$.
The table also lists $\mathrm{Ra} H/(2 \pi^2)$ for comparison with $E$.}

\label{table1}
\end{table}

Setting
$Z = \langle \frac{1}{2} |\nabla \times \nabla \times (\phi \bm{\hat z})|^2 \rangle$
and repeating the same computations as in the previous section yields bounds $E$
on the poloidal energy. Figure \ref{fig_Epol_bounds} shows the bounds as a
function of $\mathrm{Ra}$ for different cases and table \ref{table1} summarizes
the results by fitting power laws through the points at high $\mathrm{Ra}$. A
power law fit is not quite satisfactory yet at infinite $\mathrm{Pr}$ for free
slip boundaries at the $\mathrm{Ra}$ accessed in these computations.
Fig \ref{fig_Epol_fs} presents for this reason an alternative point of view in
which $E$ is directly compared with $\mathrm{Ra}H$. 

In all cases, the full optimization problem improves prefactors compared with
the direct estimate $\mathrm{Ra} H / (2\pi^2)$, but not the exponent in the
power laws or the functional form of the Rayleigh number dependence. The
prefactor improves by a factor ranging from 1.5 to 4.8.

There is an interesting curiosity about the bounds for free slip boundaries. The
bounds for infinite and general $\mathrm{Pr}$ agree to better than within $1\%$
as long as there is only one wavenumber in the active set. The difference
between the two cases becomes obvious only once the active set contains more
than one wavenumber.

Some of the fitted power laws in table \ref{table1} can be compared with data
from direct numerical simulations.
The comparison is scarce in the case of infinite $\mathrm{Pr}$ and no slip
boundaries. Numerical simulations did not go to high enough Rayleigh numbers to
be definitive about asymptotic scalings. Travis et al. \cite{Travis90} found
$\mathrm{Nu} = 0.406 \times \mathrm{Ra}^{0.284}$,
but Sotin and Labrosse \cite{Sotin99} argue that this scaling only holds for
time independent convection, and that
$\mathrm{Nu} = 0.517 \times \mathrm{Ra}^{0.269}$
once time dependence sets in. However, their simulations were restricted to
$\mathrm{Ra} < 10^7$. Theoretical arguments put forward by Grossmann and Lohse
\cite{Grossm01} lead to
$\mathrm{Nu} \propto \mathrm{Ra}^{1/3}$ and
$E_\mathrm{pol} \propto \mathrm{Ra}^{4/3}$. These exponents are the same as
those listed in table \ref{table1} for the bounds.

More results are available for infinite $\mathrm{Pr}$ and free slip boundaries.
In that case, Christensen \cite{Christ89} finds numerically 
$\mathrm{Nu} = 0.196 \times \mathrm{Ra}^{1/3}$ and Pandey et al. \cite{Pandey14}
obtain
$\mathrm{Nu} = 0.23 \times \mathrm{Ra}^{0.32}$. These relations depend to some
extent on the interval of $\mathrm{Ra}$ included in the fit (typically
$10^5-10^8$ for ref. \onlinecite{Pandey14} and up to $10^{11}$ for ref. \onlinecite{Christ89})
and also the aspect ratio of the computational volume (which was actually 2D in
Christensen's case), but the point is that there is roughly a factor of up to 2 between the bound
$H$ and the numerical results for $\mathrm{Ra}<10^{10}$ with a widening gap if
the quoted exponents persist to higher $\mathrm{Ra}$. Pandey et al.
\cite{Pandey14} also give
the kinetic energy for free slip boundaries at infinite $\mathrm{Pr}$ in a box
of size $2\sqrt(2):2\sqrt(2):1$ as 
$0.02 \times \mathrm{Ra}^{6/5}$ for $10^5 \leq \mathrm{Ra} \leq 10^8$. This is
nearly two orders of magnitude below the bound even at $\mathrm{Ra}=10^8$, and
the discrepancy is worsening with increasing $\mathrm{Ra}$ due to the
discrepancy in the exponents. The kinetic energy deduced from the characteristic
surface velocity obtained by Jarvis and Peltier \cite{Jarvis82} varies as 
$\mathrm{Ra}^{1.29}$. While closer to 1.4, this exponent is still much smaller
than the one appearing in the bound.

At finite $\mathrm{Pr}$, various simulations have found kinetic energy
approximately proportional to $\mathrm{Ra}$, so that in an all cases, the bounds
for $E_\mathrm{pol}$ severely overestimate the kinetic energy.

Another apparent overestimate is the bound for $\mathrm{Nu}$ at general
$\mathrm{Pr}$. There are arguments \cite{Kraich62} in favor of a scaling in the form 
$\mathrm{Nu} \propto \mathrm{Ra}^{1/2}$ at large $\mathrm{Ra}$, the so called
ultimate regime. However, there is as of yet no unequivocal experimental or
numerical observation of the ultimate regime and the exponents $\gamma$ in
$\mathrm{Nu} \propto \mathrm{Ra}^\gamma$ obtained from fits to observational
data are less than $1/2$. Their precise
value depends \cite{Ahlers09} on $\mathrm{Pr}$ and $\mathrm{Ra}$, but $\gamma$ is often found to
be around $1/3$. The fact that no bound better than
$\mathrm{Nu}-1 = c  \mathrm{Ra}^{1/2}$ with some constant $c$ is known for 3D
convection is sometimes seen as an indication that the ultimate regime will 
eventually be observed at high enough $\mathrm{Ra}$.

A simple argument in favor of $\mathrm{Nu} \propto \mathrm{Ra}^{1/2}$ is
dimensional analysis. This scaling must hold if the heat transport is
independent of molecular diffusivities. If one applies the same argument to
kinetic energy, one concludes that $g\alpha \Delta T h$ is the only acceptable
combination of problem parameters other than molecular diffusivities with
dimensions of velocity squared, which translates into a kinetic energy
proportional to $\mathrm{Ra}$. It is therefore very likely that the scaling in 
$\mathrm{Ra}^{3/2}$ shown by the bound will never be observed for the poloidal
kinetic energy, not even in a surmised ultimate regime. 

A scenario which generates large upper bounds is the existence of a stationary
solution with high kinetic energy which is unstable and therefore never observed
in an actual time evolution of the flow, but which prevents any bound from
approaching the observed kinetic energy of the flow. It is at present not
possible to tell whether this scenario is realized. The optimal flow field of the
SDP is not a candidate for this possible stationary flow field. The optimal flow
field contains only a few wavenumbers and not their harmonics or other mixtures,
as a solution of the Navier-Stokes equation would necessarily do.

\begin{figure}
\includegraphics[width=8cm]{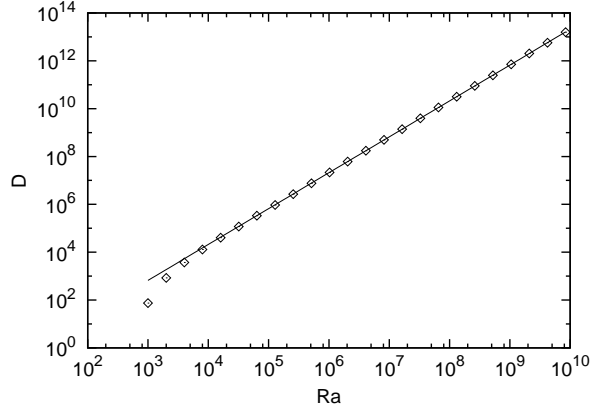}
\caption{
The optimal upper bound $D$ for the non poloidal dissipation $D_\mathrm{np}$
a function of $\mathrm{Ra}$ for general $\mathrm{Pr}$
and stress free boundaries.
The continuous line indicates the power law
$0.021 \times \mathrm{Ra}^{3/2}$.}
\label{fig_Dnp}
\end{figure}

Finally, while it is not possible to bound the energy contained in the non
poloidal components, it is possible to bound the dissipation due to them,
$D_\mathrm{np}$. Again, the energy budget immediately provides us with a bound:
\begin{equation}
D_\mathrm{np} \leq \sum_{ij} \langle \overline{(\partial_j v_i)^2} \rangle = \mathrm{Ra} (\mathrm{Nu}-1) \leq \mathrm{Ra} H
.
\end{equation}
With the data in table \ref{table1}, we conclude that
$D_\mathrm{np} \leq 0.055 \times \mathrm{Ra}^{3/2}$. Setting $Z=D_\mathrm{np}$ and
going through the whole optimization procedure leads to a better bound,
$D_\mathrm{np} \leq 0.021 \times \mathrm{Ra}^{3/2}$, shown in fig. \ref{fig_Dnp},
but the improvement is once again only in the prefactor, not in the exponent.

In an effort to improve any of these bounds, one may think of having recourse of
the maximum principle, which guarantees that, possibly after transients due to
initial conditions, the temperature
anywhere in the fluid must lie in between the temperatures of top and bottom
boundaries,
$0 \leq T=\theta+1-z \leq1$. This means that the inequality (\ref{eq:p7Mitte})
does not need to hold for any $\theta$ satisfying the boundary conditions, it is
enough if it holds for every $\theta$ satisfying the boundary conditions and the
maximum principle. 

The inequality (\ref{eq:p7discret}) decouples in $k$. It contains only quadratic
terms for $k\neq 0$. In order to violate this inequality in one of the matrix blocks
with $k \neq 0$, it is necessary that some block possesses at least one
negative eigenvalue. This condition is independent of whether the admitted
fields $\theta$ obey the maximum principle or not. For $k=0$ on the other hand,
the inequality (\ref{eq:p7discret}) or (\ref{eq:p7Mitte})
also contains linear terms and there is a non
zero $\theta$ which minimizes the left hand side of (\ref{eq:p7Mitte}). The
minimizing $\theta$ was computed for the optimal $\lambda$'s and was found to
conform to the maximum principle in all cases. It would therefore lead to no
improvement to explicitly impose this condition during the optimization process.

\section{Conclusion}

This paper presented a new method to numerically compute upper bounds for heat
transport, dissipation, and poloidal energy in Rayleigh-B\'enard convection both
for infinite and general Prandtl number. The center piece of the method is a
semidefinite program, which is numerically convenient to implement, but formally
equivalent to the background field method.

The SDP can be solved with existing numerical libraries. While it is convenient
to use an SDP solver as part of the overall procedure, it is not necessarily
efficient. The obvious alternative is to solve Euler-Lagrange equations. The
disadvantage of this option is that the Euler-Lagrange equations have many
stationary points, only one of which is the solution of the SDP which is a
convex optimization problem. On the other
hand, finding optimal bounds amounts to finding a saddle point, with a
minimization over the coefficients $\lambda$ and a maximization over the
wavenumbers in the active set. These two optimizations have to be run separately
and sequentially if an SDP solver based on central path methods
is used, but both can be combined into a
single root search if Euler-Lagrange equations are solved.

Once dissipation is bound, the mere definitions of dissipation and energy lead
to a bound on energy. Including the additional constraints employed in optimum
theory as it is used nowadays leads to tighter bounds, but only through improved
prefactors, not through smaller exponents in power laws or new functional
dependencies. It remains to be seen which additional constraints lead to further
improvement of the upper bounds. The maximum principle for the temperature field
can be of no help as it is already obeyed by the optimizing fields in the 
method presented here.

The exponent found for the bound on poloidal energy for general Prandtl numbers is not
plausible for a tight bound. Either there exist stationary unstable flows which
prevent better bounds, or the exponent can be improved
by including additional constraints derived from the Navier-Stokes
equations into the optimization process. If additional constraints have such an
effect on the bound for the poloidal energy, they may also reduce the exponent
of the known upper bound for the Nusselt number. This bound is compatible with
the expected, but as of yet unobserved, ultimate regime of convection. The
existence of this regime would be contradicted if the exponent in the bound for
the Nusselt number could be reduced.


\begin{thebibliography}{27}%
\makeatletter
\providecommand \@ifxundefined [1]{%
 \@ifx{#1\undefined}
}%
\providecommand \@ifnum [1]{%
 \ifnum #1\expandafter \@firstoftwo
 \else \expandafter \@secondoftwo
 \fi
}%
\providecommand \@ifx [1]{%
 \ifx #1\expandafter \@firstoftwo
 \else \expandafter \@secondoftwo
 \fi
}%
\providecommand \natexlab [1]{#1}%
\providecommand \enquote  [1]{``#1''}%
\providecommand \bibnamefont  [1]{#1}%
\providecommand \bibfnamefont [1]{#1}%
\providecommand \citenamefont [1]{#1}%
\providecommand \href@noop [0]{\@secondoftwo}%
\providecommand \href [0]{\begingroup \@sanitize@url \@href}%
\providecommand \@href[1]{\@@startlink{#1}\@@href}%
\providecommand \@@href[1]{\endgroup#1\@@endlink}%
\providecommand \@sanitize@url [0]{\catcode `\\12\catcode `\$12\catcode
  `\&12\catcode `\#12\catcode `\^12\catcode `\_12\catcode `\%12\relax}%
\providecommand \@@startlink[1]{}%
\providecommand \@@endlink[0]{}%
\providecommand \url  [0]{\begingroup\@sanitize@url \@url }%
\providecommand \@url [1]{\endgroup\@href {#1}{\urlprefix }}%
\providecommand \urlprefix  [0]{URL }%
\providecommand \Eprint [0]{\href }%
\providecommand \doibase [0]{http://dx.doi.org/}%
\providecommand \selectlanguage [0]{\@gobble}%
\providecommand \bibinfo  [0]{\@secondoftwo}%
\providecommand \bibfield  [0]{\@secondoftwo}%
\providecommand \translation [1]{[#1]}%
\providecommand \BibitemOpen [0]{}%
\providecommand \bibitemStop [0]{}%
\providecommand \bibitemNoStop [0]{.\EOS\space}%
\providecommand \EOS [0]{\spacefactor3000\relax}%
\providecommand \BibitemShut  [1]{\csname bibitem#1\endcsname}%
\let\auto@bib@innerbib\@empty
\bibitem [{\citenamefont {Hopf}(1941)}]{Hopf41}%
  \BibitemOpen
  \bibfield  {author} {\bibinfo {author} {\bibfnamefont {E.}~\bibnamefont
  {Hopf}},\ }\bibfield  {title} {\enquote {\bibinfo {title} {{Ein allgemeiner
  Endlichkeitssatz der Hydrodynamik}},}\ }\href@noop {} {\bibfield  {journal}
  {\bibinfo  {journal} {Mathematische Annalen}\ }\textbf {\bibinfo {volume}
  {117}},\ \bibinfo {pages} {764--775} (\bibinfo {year} {1941})}\BibitemShut
  {NoStop}%
\bibitem [{\citenamefont {Malkus}(1956)}]{Malkus56}%
  \BibitemOpen
  \bibfield  {author} {\bibinfo {author} {\bibfnamefont {W.}~\bibnamefont
  {Malkus}},\ }\bibfield  {title} {\enquote {\bibinfo {title} {Outline for a
  theory for turbulent shear flow},}\ }\href@noop {} {\bibfield  {journal}
  {\bibinfo  {journal} {J. Fluid Mech.}\ }\textbf {\bibinfo {volume} {1}},\
  \bibinfo {pages} {521--539} (\bibinfo {year} {1956})}\BibitemShut {NoStop}%
\bibitem [{\citenamefont {Howard}(1963)}]{Howard63}%
  \BibitemOpen
  \bibfield  {author} {\bibinfo {author} {\bibfnamefont {L.}~\bibnamefont
  {Howard}},\ }\bibfield  {title} {\enquote {\bibinfo {title} {Heat transport
  by turbulent convection},}\ }\href@noop {} {\bibfield  {journal} {\bibinfo
  {journal} {J. Fluid Mech.}\ }\textbf {\bibinfo {volume} {17}},\ \bibinfo
  {pages} {405--432} (\bibinfo {year} {1963})}\BibitemShut {NoStop}%
\bibitem [{\citenamefont {Busse}(1969)}]{Busse69}%
  \BibitemOpen
  \bibfield  {author} {\bibinfo {author} {\bibfnamefont {F.}~\bibnamefont
  {Busse}},\ }\bibfield  {title} {\enquote {\bibinfo {title} {{On Howard's
  upper bound for geat transport by turbulent convection}},}\ }\href@noop {}
  {\bibfield  {journal} {\bibinfo  {journal} {J. Fluid Mech.}\ }\textbf
  {\bibinfo {volume} {37}},\ \bibinfo {pages} {457--477} (\bibinfo {year}
  {1969})}\BibitemShut {NoStop}%
\bibitem [{\citenamefont {Doering}\ and\ \citenamefont
  {Constantin}(1992)}]{Doerin92}%
  \BibitemOpen
  \bibfield  {author} {\bibinfo {author} {\bibfnamefont {C.}~\bibnamefont
  {Doering}}\ and\ \bibinfo {author} {\bibfnamefont {P.}~\bibnamefont
  {Constantin}},\ }\bibfield  {title} {\enquote {\bibinfo {title} {Energy
  dissipation in shear driven turbulence},}\ }\href@noop {} {\bibfield
  {journal} {\bibinfo  {journal} {Phys. Rev. Lett.}\ }\textbf {\bibinfo
  {volume} {69}},\ \bibinfo {pages} {1648--1651} (\bibinfo {year}
  {1992})}\BibitemShut {NoStop}%
\bibitem [{\citenamefont {Doering}\ and\ \citenamefont
  {Constantin}(1996)}]{Doerin96}%
  \BibitemOpen
  \bibfield  {author} {\bibinfo {author} {\bibfnamefont {C.}~\bibnamefont
  {Doering}}\ and\ \bibinfo {author} {\bibfnamefont {P.}~\bibnamefont
  {Constantin}},\ }\bibfield  {title} {\enquote {\bibinfo {title} {{Variational
  bounds on energy dissipation in incompressible flows:III. Convection}},}\
  }\href@noop {} {\bibfield  {journal} {\bibinfo  {journal} {Phys. Rev. E}\
  }\textbf {\bibinfo {volume} {53}},\ \bibinfo {pages} {5957--5981} (\bibinfo
  {year} {1996})}\BibitemShut {NoStop}%
\bibitem [{\citenamefont {Seis}(2015)}]{Seis15}%
  \BibitemOpen
  \bibfield  {author} {\bibinfo {author} {\bibfnamefont {C.}~\bibnamefont
  {Seis}},\ }\bibfield  {title} {\enquote {\bibinfo {title} {Scaling bounds on
  dissipation in turbulent flows},}\ }\href@noop {} {\bibfield  {journal}
  {\bibinfo  {journal} {J. Fluid Mech.}\ }\textbf {\bibinfo {volume} {777}},\
  \bibinfo {pages} {591--603} (\bibinfo {year} {2015})}\BibitemShut {NoStop}%
\bibitem [{\citenamefont {Ierley}, \citenamefont {Kerswell},\ and\
  \citenamefont {Plasting}(2006)}]{Ierley06}%
  \BibitemOpen
  \bibfield  {author} {\bibinfo {author} {\bibfnamefont {G.}~\bibnamefont
  {Ierley}}, \bibinfo {author} {\bibfnamefont {R.}~\bibnamefont {Kerswell}}, \
  and\ \bibinfo {author} {\bibfnamefont {C.}~\bibnamefont {Plasting}},\
  }\bibfield  {title} {\enquote {\bibinfo {title} {{Infinite-Prandtl-number
  convection. Part 2. A singular limit of upper bound theory}},}\ }\href@noop
  {} {\bibfield  {journal} {\bibinfo  {journal} {J. Fluid Mech.}\ }\textbf
  {\bibinfo {volume} {560}},\ \bibinfo {pages} {159--227} (\bibinfo {year}
  {2006})}\BibitemShut {NoStop}%
\bibitem [{\citenamefont {Wen}\ \emph {et~al.}(2013)\citenamefont {Wen},
  \citenamefont {Chini}, \citenamefont {Dianati},\ and\ \citenamefont
  {Doering}}]{Wen13}%
  \BibitemOpen
  \bibfield  {author} {\bibinfo {author} {\bibfnamefont {B.}~\bibnamefont
  {Wen}}, \bibinfo {author} {\bibfnamefont {G.}~\bibnamefont {Chini}}, \bibinfo
  {author} {\bibfnamefont {N.}~\bibnamefont {Dianati}}, \ and\ \bibinfo
  {author} {\bibfnamefont {C.}~\bibnamefont {Doering}},\ }\bibfield  {title}
  {\enquote {\bibinfo {title} {Computational approaches to
  aspect-ratio-dependent upper bounds and heat flux in porous medium
  convection},}\ }\href@noop {} {\bibfield  {journal} {\bibinfo  {journal}
  {Phys. Lett. A}\ }\textbf {\bibinfo {volume} {377}},\ \bibinfo {pages}
  {2931--2938} (\bibinfo {year} {2013})}\BibitemShut {NoStop}%
\bibitem [{\citenamefont {Wen}\ \emph {et~al.}(2015)\citenamefont {Wen},
  \citenamefont {Chini}, \citenamefont {Kerswell},\ and\ \citenamefont
  {Doering}}]{Wen15}%
  \BibitemOpen
  \bibfield  {author} {\bibinfo {author} {\bibfnamefont {B.}~\bibnamefont
  {Wen}}, \bibinfo {author} {\bibfnamefont {G.}~\bibnamefont {Chini}}, \bibinfo
  {author} {\bibfnamefont {R.}~\bibnamefont {Kerswell}}, \ and\ \bibinfo
  {author} {\bibfnamefont {C.}~\bibnamefont {Doering}},\ }\bibfield  {title}
  {\enquote {\bibinfo {title} {{Time-stepping approach for solving upper-bound
  problems: Application to two-dimensional Rayleigh-B\'enard convection}},}\
  }\href@noop {} {\bibfield  {journal} {\bibinfo  {journal} {Phys. Rev. E}\
  }\textbf {\bibinfo {volume} {92}},\ \bibinfo {pages} {043012} (\bibinfo
  {year} {2015})}\BibitemShut {NoStop}%
\bibitem [{\citenamefont {Fantuzzi}\ and\ \citenamefont
  {Wynn}(2016)}]{Fantuz16}%
  \BibitemOpen
  \bibfield  {author} {\bibinfo {author} {\bibfnamefont {G.}~\bibnamefont
  {Fantuzzi}}\ and\ \bibinfo {author} {\bibfnamefont {A.}~\bibnamefont
  {Wynn}},\ }\bibfield  {title} {\enquote {\bibinfo {title} {{Optimal bounds
  with semidefinite programming: An application to stress-driven shear
  flows}},}\ }\href@noop {} {\bibfield  {journal} {\bibinfo  {journal} {Phys.
  Rev. E}\ }\textbf {\bibinfo {volume} {93}},\ \bibinfo {pages} {043308}
  (\bibinfo {year} {2016})}\BibitemShut {NoStop}%
\bibitem [{\citenamefont {Boyd}\ and\ \citenamefont
  {Vandenberghe}(2004)}]{Boyd04}%
  \BibitemOpen
  \bibfield  {author} {\bibinfo {author} {\bibfnamefont {S.}~\bibnamefont
  {Boyd}}\ and\ \bibinfo {author} {\bibfnamefont {L.}~\bibnamefont
  {Vandenberghe}},\ }\href@noop {} {\emph {\bibinfo {title} {{Convex
  Optimization}}}}\ (\bibinfo  {publisher} {Cambridge University Press},\
  \bibinfo {address} {Cambridge},\ \bibinfo {year} {2004})\BibitemShut
  {NoStop}%
\bibitem [{\citenamefont {Tobasco}, \citenamefont {Goluskin},\ and\
  \citenamefont {Doering}(2017)}]{Tobasc18}%
  \BibitemOpen
  \bibfield  {author} {\bibinfo {author} {\bibfnamefont {I.}~\bibnamefont
  {Tobasco}}, \bibinfo {author} {\bibfnamefont {D.}~\bibnamefont {Goluskin}}, \
  and\ \bibinfo {author} {\bibfnamefont {C.}~\bibnamefont {Doering}},\
  }\bibfield  {title} {\enquote {\bibinfo {title} {Optimal bounds and extremal
  trajectories for time averages in dynamical systems},}\ }\href@noop {}
  {\bibfield  {journal} {\bibinfo  {journal} {arXiv:1705.07096}\ } (\bibinfo
  {year} {2017})}\BibitemShut {NoStop}%
\bibitem [{\citenamefont {Goluskin}(2017)}]{Golusk18}%
  \BibitemOpen
  \bibfield  {author} {\bibinfo {author} {\bibfnamefont {D.}~\bibnamefont
  {Goluskin}},\ }\bibfield  {title} {\enquote {\bibinfo {title} {{Bounding
  averages rigorously using semidefinite programming: mean moments of the
  Lorenz system}},}\ }\href@noop {} {\bibfield  {journal} {\bibinfo  {journal}
  {arXiv:1610.05335}\ } (\bibinfo {year} {2017})}\BibitemShut {NoStop}%
\bibitem [{\citenamefont {Whithehead}\ and\ \citenamefont
  {Doering}(2012)}]{Whiteh12}%
  \BibitemOpen
  \bibfield  {author} {\bibinfo {author} {\bibfnamefont {J.}~\bibnamefont
  {Whithehead}}\ and\ \bibinfo {author} {\bibfnamefont {C.}~\bibnamefont
  {Doering}},\ }\bibfield  {title} {\enquote {\bibinfo {title} {Rigid bounds on
  heat transport by a fluid between slippery boundaries},}\ }\href@noop {}
  {\bibfield  {journal} {\bibinfo  {journal} {J. Fluid Mech.}\ }\textbf
  {\bibinfo {volume} {707}},\ \bibinfo {pages} {241--259} (\bibinfo {year}
  {2012})}\BibitemShut {NoStop}%
\bibitem [{\citenamefont {Nobili}\ and\ \citenamefont {Otto}(2017)}]{Nobili17}%
  \BibitemOpen
  \bibfield  {author} {\bibinfo {author} {\bibfnamefont {C.}~\bibnamefont
  {Nobili}}\ and\ \bibinfo {author} {\bibfnamefont {F.}~\bibnamefont {Otto}},\
  }\bibfield  {title} {\enquote {\bibinfo {title} {{Limitations of the
  background field method applied to Rayleigh-B\'enard convection}},}\
  }\href@noop {} {\bibfield  {journal} {\bibinfo  {journal} {J. Math. Phys.}\
  }\textbf {\bibinfo {volume} {58}},\ \bibinfo {pages} {093102} (\bibinfo
  {year} {2017})}\BibitemShut {NoStop}%
\bibitem [{\citenamefont {Vitanov}(1998{\natexlab{a}})}]{Vitano98}%
  \BibitemOpen
  \bibfield  {author} {\bibinfo {author} {\bibfnamefont {N.}~\bibnamefont
  {Vitanov}},\ }\bibfield  {title} {\enquote {\bibinfo {title} {{Upper bound on
  the heat transport in a horizontal fluid layer of infinite Prandtl
  number}},}\ }\href@noop {} {\bibfield  {journal} {\bibinfo  {journal} {Phys.
  Lett. A}\ }\textbf {\bibinfo {volume} {248}},\ \bibinfo {pages} {338--346}
  (\bibinfo {year} {1998}{\natexlab{a}})}\BibitemShut {NoStop}%
\bibitem [{\citenamefont {Vitanov}(1998{\natexlab{b}})}]{Vitano98b}%
  \BibitemOpen
  \bibfield  {author} {\bibinfo {author} {\bibfnamefont {N.}~\bibnamefont
  {Vitanov}},\ }\bibfield  {title} {\enquote {\bibinfo {title} {{On the Optimum
  Theory of Turbulence}},}\ }\href@noop {} {\bibfield  {journal} {\bibinfo
  {journal} {doctoral thesis, University of Bayreuth}\ } (\bibinfo {year}
  {1998}{\natexlab{b}})}\BibitemShut {NoStop}%
\bibitem [{\citenamefont {Schmitt}\ and\ \citenamefont {von
  Wahl}(1992)}]{Schmit92}%
  \BibitemOpen
  \bibfield  {author} {\bibinfo {author} {\bibfnamefont {B.~J.}\ \bibnamefont
  {Schmitt}}\ and\ \bibinfo {author} {\bibfnamefont {W.}~\bibnamefont {von
  Wahl}},\ }\bibfield  {title} {\enquote {\bibinfo {title} {Decomposition into
  poloidal fields, toroidal fields, and mean flow},}\ }\href@noop {} {\bibfield
   {journal} {\bibinfo  {journal} {Differential and Integral Equations}\
  }\textbf {\bibinfo {volume} {5}},\ \bibinfo {pages} {1275--1306} (\bibinfo
  {year} {1992})}\BibitemShut {NoStop}%
\bibitem [{\citenamefont {Travis}, \citenamefont {Olson},\ and\ \citenamefont
  {Schubert}(1990)}]{Travis90}%
  \BibitemOpen
  \bibfield  {author} {\bibinfo {author} {\bibfnamefont {B.}~\bibnamefont
  {Travis}}, \bibinfo {author} {\bibfnamefont {P.}~\bibnamefont {Olson}}, \
  and\ \bibinfo {author} {\bibfnamefont {G.}~\bibnamefont {Schubert}},\
  }\bibfield  {title} {\enquote {\bibinfo {title} {{The transition from
  two-dimensional to three-dimensional planforms in infinite Prandtl-number
  thermal convection}},}\ }\href@noop {} {\bibfield  {journal} {\bibinfo
  {journal} {J. Fluid Mech.}\ }\textbf {\bibinfo {volume} {216}},\ \bibinfo
  {pages} {71--91} (\bibinfo {year} {1990})}\BibitemShut {NoStop}%
\bibitem [{\citenamefont {Sotin}\ and\ \citenamefont
  {Labrosse}(1999)}]{Sotin99}%
  \BibitemOpen
  \bibfield  {author} {\bibinfo {author} {\bibfnamefont {C.}~\bibnamefont
  {Sotin}}\ and\ \bibinfo {author} {\bibfnamefont {S.}~\bibnamefont
  {Labrosse}},\ }\bibfield  {title} {\enquote {\bibinfo {title}
  {{Three-dimensional thermal convection in an iso-viscous, infinite Prandtl
  number fluid heated from within and from below: applications to the heat
  transfer of planetary mantles}},}\ }\href@noop {} {\bibfield  {journal}
  {\bibinfo  {journal} {Phys. Earth Planet. Inter.}\ }\textbf {\bibinfo
  {volume} {112}},\ \bibinfo {pages} {171--190} (\bibinfo {year}
  {1999})}\BibitemShut {NoStop}%
\bibitem [{\citenamefont {Grossmann}\ and\ \citenamefont
  {Lohse}(2001)}]{Grossm01}%
  \BibitemOpen
  \bibfield  {author} {\bibinfo {author} {\bibfnamefont {S.}~\bibnamefont
  {Grossmann}}\ and\ \bibinfo {author} {\bibfnamefont {D.}~\bibnamefont
  {Lohse}},\ }\bibfield  {title} {\enquote {\bibinfo {title} {{Thermal
  Convection at Large Prandtl Numbers}},}\ }\href@noop {} {\bibfield  {journal}
  {\bibinfo  {journal} {Phys. Rev. Lett.}\ }\textbf {\bibinfo {volume} {86}},\
  \bibinfo {pages} {3316--3319} (\bibinfo {year} {2001})}\BibitemShut {NoStop}%
\bibitem [{\citenamefont {Christensen}(1989)}]{Christ89}%
  \BibitemOpen
  \bibfield  {author} {\bibinfo {author} {\bibfnamefont {U.}~\bibnamefont
  {Christensen}},\ }\bibfield  {title} {\enquote {\bibinfo {title} {{The heat
  transport by convection rolls with free boundaries at high Rayleigh
  number}},}\ }\href@noop {} {\bibfield  {journal} {\bibinfo  {journal}
  {Geophys. Astrophys. Fluid Dyn.}\ }\textbf {\bibinfo {volume} {46}},\
  \bibinfo {pages} {93--103} (\bibinfo {year} {1989})}\BibitemShut {NoStop}%
\bibitem [{\citenamefont {Pandey}, \citenamefont {Verma},\ and\ \citenamefont
  {Mishra}(2014)}]{Pandey14}%
  \BibitemOpen
  \bibfield  {author} {\bibinfo {author} {\bibfnamefont {A.}~\bibnamefont
  {Pandey}}, \bibinfo {author} {\bibfnamefont {M.}~\bibnamefont {Verma}}, \
  and\ \bibinfo {author} {\bibfnamefont {P.}~\bibnamefont {Mishra}},\
  }\bibfield  {title} {\enquote {\bibinfo {title} {{Scaling of heat flux and
  energy spectrum for very large Prandtl number convection}},}\ }\href@noop {}
  {\bibfield  {journal} {\bibinfo  {journal} {Phys. Rev. E}\ }\textbf {\bibinfo
  {volume} {89}},\ \bibinfo {pages} {023006} (\bibinfo {year}
  {2014})}\BibitemShut {NoStop}%
\bibitem [{\citenamefont {Jarvis}\ and\ \citenamefont
  {Peltier}(1982)}]{Jarvis82}%
  \BibitemOpen
  \bibfield  {author} {\bibinfo {author} {\bibfnamefont {G.}~\bibnamefont
  {Jarvis}}\ and\ \bibinfo {author} {\bibfnamefont {W.}~\bibnamefont
  {Peltier}},\ }\bibfield  {title} {\enquote {\bibinfo {title} {Mantle
  convection as a boundary layer phenomenon},}\ }\href@noop {} {\bibfield
  {journal} {\bibinfo  {journal} {Geophys. J. Roy. Astron. Soc.}\ }\textbf
  {\bibinfo {volume} {68}},\ \bibinfo {pages} {389--427} (\bibinfo {year}
  {1982})}\BibitemShut {NoStop}%
\bibitem [{\citenamefont {Kraichnan}(1962)}]{Kraich62}%
  \BibitemOpen
  \bibfield  {author} {\bibinfo {author} {\bibfnamefont {R.}~\bibnamefont
  {Kraichnan}},\ }\bibfield  {title} {\enquote {\bibinfo {title} {{Turbulent
  thermal convection at arbitrary Prandtl number}},}\ }\href@noop {} {\bibfield
   {journal} {\bibinfo  {journal} {Phys. Fluids}\ }\textbf {\bibinfo {volume}
  {5}},\ \bibinfo {pages} {1374--1389} (\bibinfo {year} {1962})}\BibitemShut
  {NoStop}%
\bibitem [{\citenamefont {Ahlers}, \citenamefont {Grossmann},\ and\
  \citenamefont {Lohse}(2009)}]{Ahlers09}%
  \BibitemOpen
  \bibfield  {author} {\bibinfo {author} {\bibfnamefont {G.}~\bibnamefont
  {Ahlers}}, \bibinfo {author} {\bibfnamefont {S.}~\bibnamefont {Grossmann}}, \
  and\ \bibinfo {author} {\bibfnamefont {D.}~\bibnamefont {Lohse}},\ }\bibfield
   {title} {\enquote {\bibinfo {title} {{Heat transfer \& large-scale dynamics
  in turbulent Rayleigh-B\'enard convection}},}\ }\href@noop {} {\bibfield
  {journal} {\bibinfo  {journal} {Rev. Mod. Phys.}\ }\textbf {\bibinfo {volume}
  {81}},\ \bibinfo {pages} {503--537} (\bibinfo {year} {2009})}\BibitemShut
  {NoStop}%
\end{thebibliography}

%

\end{document}